\documentclass[a4paper, oneside, twocolumn, notitlepage, 10pt]{extarticle_ecoc}
\usepackage{ecoc}
\usepackage{booktabs}
\usepackage{setspace}
\usepackage{color}

\newcommand{\comm}[1]{}
\newcommand{\imag}{\jmath}
\newcommand{\vect}[1]{\boldsymbol{#1}}

\begin{document}
\selectlanguage{american}    % Standard Language

%-------------------------------------------------- Title -----------------------------------------------------%

% CH: just a suggestion:
\title{ASIC Implementation of Time-Domain Digital Backpropagation with
Deep-Learned Chromatic Dispersion Filters}%

%------------------------------------------------- Authors-----------------------------------------------------%

\author{
  Christoffer Fougstedt\textsuperscript{(1)}, Christian
	H\"ager\textsuperscript{(2,3)}, Lars Svensson\textsuperscript{(1)},\\
	Henry D. Pfister\textsuperscript{(3)}, and Per Larsson-Edefors\textsuperscript{(1)}}

\maketitle                  % Create title and author

%------------------------------------------ Description of Authors ----------------------------------------------%

\begin{strip}
  \begin{author_descr}

   \textsuperscript{(1)} Department of Computer Science and Engineering,
	 Chalmers University of Technology, Sweden\\
	 \textsuperscript{(2)} Department of Electrical Engineering,
	 Chalmers University of Technology, Sweden\\
   \textsuperscript{(3)} Department of Electrical and Computer Engineering, Duke University, USA

   \uline{Email: chrfou@chalmers.se}
 \end{author_descr}
\end{strip}

\setstretch{1.1}

%-------------------------------------------------- Abstract ---------------------------------------------------------%

\begin{strip}
  \begin{ecoc_abstract}
		% NOTE: Don't use a blank line here but start abstract right away
		% to avoid an extra line break
        We consider time-domain digital backpropagation with
        chromatic dispersion filters jointly optimized and quantized
        using machine-learning techniques. Compared
        to the baseline implementations, we show improved BER
        performance and \textgreater 40\% power dissipation reductions in 28-nm CMOS.

		%We consider time-domain digital backpropagation based on
		%chromatic dispersion filters that are jointly optimized and
		%quantized using machine-learning techniques. We evaluate the
		%fixed-point requirements, power dissipation, and chip area in
		%$28$-nm CMOS.

%		We characterize the fixed-point requirements, power dissipation,
%		and chip area of time-domain digital backpropagation, \blue{using
%		quantized chromatic-dispersion filters jointly-optimized using
%		machine-learning techniques,} in $28$-nm CMOS.
		
		%assuming that the chromatic dispersion filters in all steps are
		%jointly optimized using machine learning techniques.
  \end{ecoc_abstract}
\end{strip}

\section{Introduction}

%\begin{figure}[t]
%   \centering
%        \includegraphics[height=3.9cm]{../figs/float_sig.pdf}
%        \caption{BER as a function of 100-km spans for 7--10-bit coefficients,
%          and the frequency-domain (FD) reference algorithm.
%          Performance for 9- and 10-bit pairs are indistinguishable at this scale.}
%    \label{fig:float_sig}
%\end{figure}

%\red{General DBP/NL compensation intro with related work}

%,xiao2017low

Fiber nonlinearities impose a fundamental limitation on transmission
performance and various nonlinear compensation schemes have been
proposed. Our focus is on digital backpropagation (DBP) which emulates
backward fiber propagation using digital signal processing (DSP).
Different optimizations of DBP algorithms have been
studied~\cite{ip2008jlt,yan_ECOC11,rafique_OE11,napoli_JLT14} but only
recently have DSP hardware implementation aspects been
considered~\cite{fougstedt+:ofc17,fougstedt+:ecoc17,Martins2018,Sherborne2018}.

A major issue with DBP based on the split-step Fourier method (SSFM)
is the large complexity caused by the fast Fourier transforms (FFTs).
Time-domain DBP (TD-DBP) with finite impulse response (FIR) filters
may be competitive\cite{Zhu2009, Goldfarb2009,
fougstedt+:ofc17,fougstedt+:ecoc17, hager+:ofc2018, Haeger2018isit,
Martins2018}, assuming that the chromatic dispersion (CD) steps are
sufficiently short. Design methods for the FIR filters include least
squares\cite{Sheikh2016,fougstedt+:ofc17,fougstedt+:ecoc17} or
wavelets\cite{Goldfarb2009}, but accumulating truncation errors due to
repeated filter use can lead to severe performance degradations.
Ideally, the coefficients of all filters in the entire DBP algorithm
should be optimized jointly. It has recently been shown that this can
be accomplished in an efficient way using deep learning, leading to
very short CD filters per step \cite{hager+:ofc2018, Haeger2018isit}.

In this paper, we study TD-DBP based on deep-learned CD filters from
an ASIC implementation perspective. In particular, we evaluate the
finite-resolution requirements in terms of the minimum number of
quantization bits for the filter coefficients and signal. Moreover,
hardware synthesis results for power dissipation and chip area in
$28$-nm CMOS are presented and discussed.

% CH: I moved this to the conclusion:
%
%It is shown that the obtained filters have similar signal resolution
%requirements compared to our previous results
%in\cite{fougstedt+:ofc17,fougstedt+:ecoc17}, and significantly reduced
%coefficient resolution requirements. Moreover, reduced filter lengths
%translate into lower power dissipation and chip area.

\section{Time-Domain Digital Backpropagation}

%\begin{align*}
%	\frac{\partial A}{\partial z} = (\hat{D} + \hat{N}) A = \left(
%	- \imag \frac{\beta_2}{2} \frac{\partial^2}{\partial t^2}
% 	- \frac{\alpha}{2}
%	\right)
%	A + \imag \gamma |A|^2 A
%\end{align*}

%DBP refers to solving the nonlinear Schr\"odinger equation (NLSE) with
%negated fiber parameters.

Light propagation in an optical fiber is described by the nonlinear
Schr\"odinger equation (NLSE). In general, the NLSE needs to be solved
using numerical methods, where, in the context of DBP, the
SSFM~\cite{ip2008jlt,Agrawal2013} is the most prominent one. The SSFM
divides the transmission distance into $M$ steps of size
$\delta_\ell$, $\ell = 1, \dots, M$. The solution for step $\ell$ is
then approximated by applying a linear filtering step with frequency
response $H_\ell(\omega) = e^{\imag \frac{\beta_2}{2} \delta_{\ell}
\omega^2}$, where $\beta_2$ is the CD coefficient, and a nonlinear
phase rotation step $\rho_\ell(x) = x e^{- \imag \gamma \delta_\ell
|x|^2}$, where $\gamma$ is the Kerr parameter.

%and {\color{red} previous work\cite{fougstedt}}

In TD-DBP, the linear step is implemented as a direct convolution of
the signal with a symmetric FIR filter $H^{(\ell)}(z)$. This can be
more efficient than FFT-based filtering, where the efficiency
crossover point depends on implementation details. Assuming similar
fixed-point requirements, a first-order estimate based on our hardware
assumptions is between $25$--$30$ filter taps and similar values can
be found in the literature\cite{Borgerding2006, Fougstedt2018c}. However, taking into
account the potentially increased fixed-point requirements for
multiple cascaded FFT/IFFTs\cite{Sherborne2018} as employed in
frequency-domain split-step DBP, the efficiency crossover point may be
even higher. 

%this does not take into account fixed-point requirements which are,
%to the best of our knowledge, unknown for multiple cascaded FFT/IFFTs
%as employed in frequency-domain split-step DBP.

%{\color{red} However, this does not take into account the increase
%finite-precision accuracy due to multiple cascaded FFTs and IFFTs in
%frequency-domain split-step DBP.  An additional benefit of time-domain
%techniques is that they allow for joint-optimization of the quantized
%impulse-responses, allowing for partial cancellation of
%quantization-induced frequency-response
%errors~\cite{fougstedt+:ecoc17}, and thus low word length
%requirements.}

%``On a fixed-point DSP with a single-cycle multiply-accumulate
%instruction, the crossover value can be over 50 taps''
%\cite{Borgerding2006}.

\section{Joint Filter Optimization using Deep Learning}

\begin{figure*}
	\centering
		\includegraphics{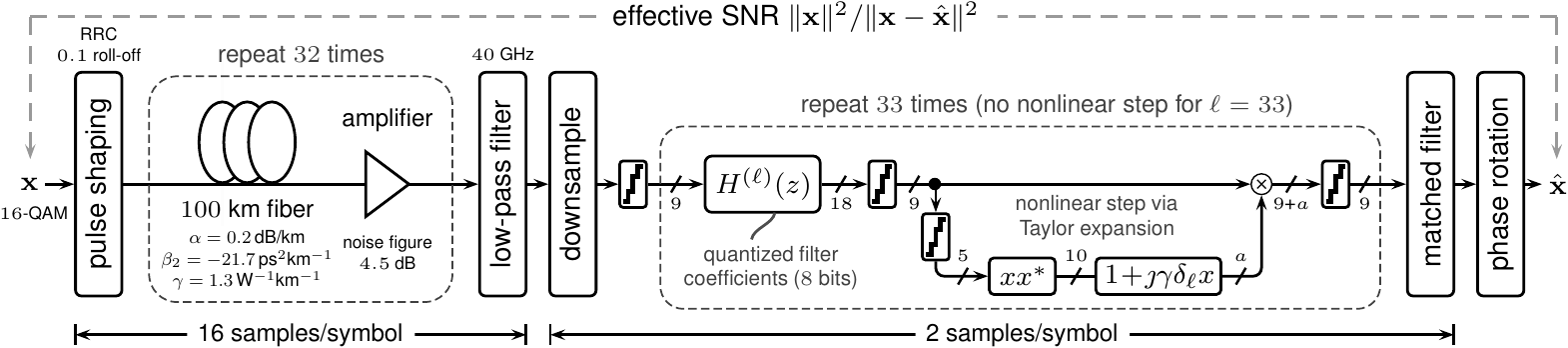}

	\vspace*{0.25cm} \caption{System model assuming TD-DBP based on the
	symmetric SSFM with $1$ StPS and $9$-bit signal
	quantization. Fixed-point scaling factors are implicitly propagated, and rounded
	to the closest power-of-two to allow for hardware-efficient
	implementation.\vspace*{0.30cm}}
	\label{fig:system_model}
\end{figure*}

%

%, i.e., $2M(K+1)$ real optimization parameters in total,

The system setup is shown in Fig.~\ref{fig:system_model}, where
the four quantization blocks can be ignored for now.
TD-DBP\cite{fougstedt+:ofc17,fougstedt+:ecoc17} is based on a
$1$-step-per-span (StPS) symmetric SSFM\cite{Agrawal2013} with
simplified ``hardware-friendly'' nonlinear steps according to a
first-order Taylor expansion $\rho_\ell(x) \approx x (1 + \imag
\gamma \delta_\ell |x|^2)$.

We use $\vect{h}^{(\ell)} = (h_{-K}^{(\ell)}, \dots, h_0, \dots,
h_{K}^{(\ell)})$ to denote the coefficients of $H^{(\ell)}(z)$, where
all $M=33$ filters have $T = 2K+1$ taps. The coefficients
$\vect{h}^{(\ell)}$ are typically optimized separately for each step,
e.g., by approximating $H_\ell(\omega)$ via least squares. A different
approach is to perform a joint optimization of all coefficients
$\theta = \{\vect{h}^{(1)}, \dots, \vect{h}^{(M)}\}$ based on a
suitable system criterion. Here, deep learning via stochastic gradient
descent is used, similar to\cite{hager+:ofc2018, Haeger2018isit}. To
that end, the system in Fig.~\ref{fig:system_model} is implemented in
TensorFlow and the effective SNR is used as the optimization
criterion.

\begin{figure}%[!b]
  \centering
  \includegraphics[width=1\columnwidth]{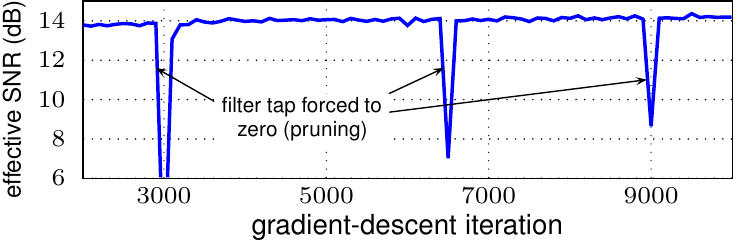}

	\vspace*{0.05cm}
	\caption{Typical learning curve when pruning filter taps.\vspace*{-0.00cm}}
	\label{fig:learning_curve}
\end{figure}

Compared to \cite{hager+:ofc2018, Haeger2018isit}, a slightly
different optimization procedure is employed as follows.  All filters
are initialized with constrained least-squares\cite{Sheikh2016}
(LS-CO) coefficients, with filter length $T' > T$ chosen large enough
to ensure good performance. Then, the filters are successively pruned
down to their target length $T$ by forcing the corresponding outermost
taps to zero at certain iterations in the gradient-descent
optimization. A typical learning curve is shown in
Fig.~\ref{fig:learning_curve}. While the instantaneous SNR loss due to
pruning can be large, gradient descent quickly recovers. We found that
the quality of the final filters is relatively insensitive to the
pruning details (e.g., which filter tap is pruned in which iteration),
as long as the pruning steps are sufficiently spread out and $T$ is
not too small.  \vspace*{-0.15cm}
%In total, $M(T'-T)/2$ pruning steps occur and

%, where the dips correspond to performing pruning

\section{Filter Coefficient and Signal Quantization}

The filter optimization assumes floating-point coefficients whereas
quantized coefficients are required for the ASIC
implementation. We make use of TensorFlow's fixed-point operations
which allow for a joint optimization of the quantized impulse
responses. This approach results in a partial cancellation of
quantization-induced frequency-response errors. All steps are jointly
optimized, thus allowing for improved cancellation in comparison to
pair-wise optimization~\cite{fougstedt+:ecoc17}. In particular, to
find good quantized filter coefficients, TensorFlow's ``fake quantization'' operations
are applied to the coefficient variables which are activated after the
(floating-point) optimization has converged. For fake quantizations,
the gradient computation and parameter updates are still performed in
floating point, allowing us to continue the training for a few more
iterations ($500$--$1000$) with a very small learning rate. We found
that this approach results in close-to
floating-point performance at very low coefficient word lengths.

The locations where signal quantization occurs in our hardware model is
indicated by the quantization blocks in Fig.~\ref{fig:system_model}.
Note that, since the resulting output word length of each
multiplication is the sum of the lengths of the operands, output
rounding is required. DBP is sensitive to bias in rounding due to the
many cascaded steps, and truncation thus imparts a penalty. A
better performing, low-complexity option is to add $0.5$ unit of least
precision (in regards to the target word length) before truncation.
This gives a close-to unbiased rounding, with $0.5$ always rounding
up. We remark that signal quantization is not implemented in
TensorFlow for the filter optimization.

%and hence the resulting quantization noise is not
%taken into account for the joint filter optimization.

%in learned we optimize each coefficient stand-alone by adjusting the
%amplitude (and thus how the dynamic range is used). This set of
%coefficients are convoluted with the used RRC pulse, and the quantized
%coefficient with the mean squared error from the floating-point case
%is selected.

%Rounding
%\begin{align}
%	Q(x) = \text{sgn}(x) \lfloor \frac{|x|}{\Delta} + 0.5 \rfloor \Delta
%\end{align}

\section{ASIC Implementation}

%, using the LS-CO and learned filter coefficients

We implemented a $96$-parallel TD-DBP, operating at $416.7$\,MHz. Each
step consists of a reconfigurable parallel FIR filter, exploiting tap symmetry
to reduce the number of multiplications.
Due to the Gaussian-like statistics of the signal, clipping is
performed, if necessary, to effectively use the dynamic
range. Numerical accuracy was verified with respect to the reference MATLAB step.

The implemented TD-DBP step was synthesized using Cadence Genus and a
low-power $28$-nm CMOS library, characterized at the slow process
corner at $125^\circ$\,C, and a supply voltage of $0.6$ V. In order to
generate accurate internal circuit switching statistics, the
implemented netlist was then simulated using input data generated in
the system model. Internal node switching activity was saved, back-annotated to the netlist, and power
was estimated using Cadence Genus at the typical process corner, at
$25^\circ$\,C, averaging over four different impulse responses.

%\vspace*{-0.2cm}
\section{Results and Discussion}

%Various methods have been proposed to quantize the CD filter
%coefficients in TD-DBP\cite{fougstedt+:ofc17, fougstedt+:ecoc17,
%Martins2018}.
%\vspace*{-0.07cm}

\begin{figure}[!t]
	\centering
		\includegraphics{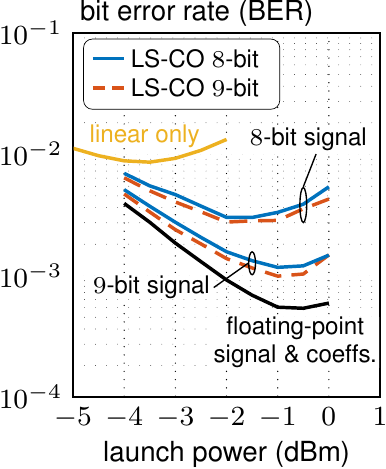}
		%$\,$
		\includegraphics{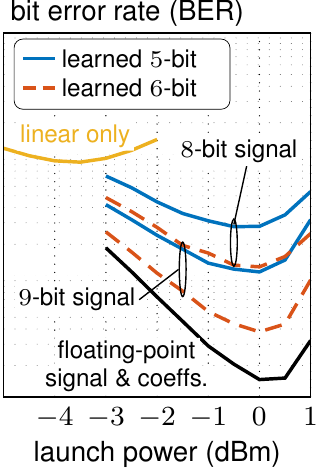}

	\vspace*{0.05cm}
	\caption{Results for 25-tap LS-CO and
	15-tap learned filters.}
%	\caption{Simulation results for 25-tap LS-CO (left) and
%	15-tap deep-learned CD filter coefficients (right).}
	\label{fig:ber}
	
\end{figure}

As a baseline, LS-CO filters are used in all steps\cite{fougstedt+:ofc17,fougstedt+:ecoc17}, in which case $25$
taps are required for good performance. Deep learning and pruning
reduce the filter length from $25$ to $15$, while improving
floating-point performance, as shown in Fig.~\ref{fig:ber}. For the
fixed-point implementation, $8$- and $9$-bit signals are considered.
Co-optimized quantization\cite{fougstedt+:ecoc17} is used for the
baseline filters, resulting in good performance for $8$- and $9$-bit
coefficients. The learned filters have comparable signal resolution
requirements (i.e., $8$--$9$ bits), but much lower
coefficient-resolution requirements. In particular, joint optimization
gives negligible penalty using $6$-bit taps compared to floating-point
and acceptable performance is achievable using $5$-bit taps.

TD-DBP with, on average, $4$ bits per tap has been
shown\cite{Martins2018}. These results rely on the specific
FIR filter shape caused by a direct truncation of the inverse CD
frequency response. Unfortunately, the resulting filter is very long, requiring $301$ symmetric taps in the considered system
\cite{Martins2018}.

% Martins are a bit unclear with their #taps, it is actually 301
% symmetric according to their figure, giving the same #mults as 151
% non-sym.

%\newpage

%$2^{21}$ symbols, were forward-propagated using 16 SaPS. The output
%was brickwall filtered and down-sampled to 2 SaPS.

%\begin{figure}[h!]
%  \centering
%  \includegraphics[width=0.48\textwidth]{./figures/ldbp.pdf}
%  \caption{\red{LDBP performance, note X-axis.} \label{fig:ldbpp}}
%\end{figure}
%\begin{figure}[h!]
%  \centering
%  \includegraphics[width=0.48\textwidth]{./figures/tddbp.pdf}
%  \caption{\red{TD-DBP performance, note X-axis.} \label{fig:ldbpp}}
%\end{figure}
%\cite{Ranzini2015} estimates chip area based on 256-point FFTs? From
%what I understand, they get $150$ mm$^{2}$ for 1 StPS 25 spans and pay
%no attenation to fixed-point issues.

%In order to put the results in
%Tab.~\ref{tab:powar2} into perspective,

The hardware synthesis results are shown in Tab.~\ref{tab:powar2},
with averaging over four different impulse responses for the learned
coefficients. The maximum deviation from the average was found to be
$2\%$. Fewer taps and lower coefficient resolution for the
deep-learned filters translate into sizable reductions in terms of
power dissipation and chip area.  We also compare to the (few)
implementation results for linear CD compensation available in the
literature. Pillai et al.\cite{Pillai2014} estimated the power
dissipation of CD compensation for $2400$-km propagation at $94$
pJ/bit (or $113$ pJ/information bit at a code overhead of $20$\%) in
$28$-nm CMOS.  An actual $40$-nm ASIC receiver implementation showed
$221$ pJ/bit for CD compensation of $3500$-km
fiber\cite{Crivelli2014}, which translates to roughly $150$ pJ/bit in
a $28$-nm process technology. In our case, the $6$-bit learned
coefficients with a $9$-bit signal resolution result in
$33\textnormal{~steps}\times 0.2\textnormal{~W} /
80\textnormal{~Gb/s}=83$ pJ/bit for $3200$-km transmission. While such
comparisons are not perfectly fair, they show that TD-DBP and deep
learning offer a viable route to implementation of nonlinearity
compensation based on split-step methods.

%\vspace*{-0.1cm}

\section{Conclusion}

We have studied TD-DBP based on deep-learned CD filters from an ASIC
perspective. It was shown that the obtained filters have similar
signal resolution requirements compared to our previous
work\cite{fougstedt+:ofc17,fougstedt+:ecoc17}, and significantly
reduced coefficient resolution requirements. Moreover, reduced filter
lengths directly translate into lower power dissipation and chip area.
Compared to LS-CO-based TD-DBP, a power-dissipation reduction of
\textgreater${40}\%$ for all considered configurations is shown.

%\vspace*{-0.3cm}
%\section{Acknowledgement}

\vspace*{0.2cm}

{\footnotesize
\begin{spacing}{0.9}

{\it Acknowledgement:}
This work was financially supported by the Knut and Alice Wallenberg
Foundation. This work is also part of a project that has received
funding from the European Union's Horizon 2020 research and innovation
programme under the Marie Sk\l{}odowska-Curie grant agreement
No.~749798.
\vspace*{-0.35cm}
\end{spacing}
}

\newcommand{\NegSp}{$\!\!\!$}
\newcommand{\NegSpT}{$\!\!$}

\begin{table}[t!]
	\footnotesize
\centering
\caption{$0.6$-V $96$-parallel results for power ($P$) and area ($A$) per
TD-DBP step ($100$ km, $20$ Gbaud, single polarization).}
\renewcommand{\arraystretch}{0.7}
  \setlength{\tabcolsep}{4pt}
\label{tab:powar2}
\begin{tabular}{@{}c|c|cc|cc@{}}
\toprule
 coeffs.~\& & filter & \multicolumn{2}{c|}{$8$-bit signal} &
 \multicolumn{2}{c}{$9$-bit signal} \\
 word length & taps & $\!\!$ $P$ (W)&\NegSp $A$ (mm$^2$) $\!\!$& $\!\!$ $P$ (W)&\NegSp $A$ (mm$^2$) \\ \midrule
 LS-CO $8$-bit & $25$ & $0.28$  &\NegSp $1.21$ $\,$& $0.31$ &\NegSp $1.30$ $\,$ \\
 LS-CO $9$-bit & $25$ & $0.34$  &\NegSp $1.38$ $\,$& $0.37$ &\NegSp $1.54$ $\,$\\ \midrule
 learned $5$-bit & $15$ & $0.15$  &\NegSp $0.61$ $\,$& $0.18$ &\NegSp $0.69$ $\,$\\
 learned $6$-bit & $15$ & $0.17$  &\NegSp $0.69$ $\,$& $0.20$ &\NegSp $0.81$ $\,$\\ \bottomrule
\end{tabular}
\vspace*{-0.35cm}
\end{table}

%\bibliographystyle{osajnl}
%\begin{spacing}{1.0}
%\bibliography{references,references_haeger}

\renewcommand\baselinestretch{0.97}

% Swap out above line and insert .bbl later
%\comm{
%\begin{thebibliography}{1}
%\bibitem{ref1}
%F. M. Lastname et al., ``\uline{Full Paper Titles are Mandatory for Adequate Referencing in Web Search Engines},'' J. Lightwave Technol., Vol. {\bf 12}, no. 5, p. 456 (1990).
%\bibitem{ref2}
%F. M. Lastname et al., ``Why Should we Use Photonics?,'' Proc. OFC, WeA3, Anaheim (2005).
%\bibitem{ref3}
%F. M. Lastname et al., ``Connecting the Dots,'' Proc. ECOC, Tu.9.G.1, London (2005).
%\bibitem{ref4}
%F. M. Lastname et al., ``Why Connect the Dots?,'' Photon. Technol. Lett., Vol. {\bf 12}, no. 3, p. 34 (2001).
%\bibitem{ref5}
%F. M. Lastname et al., ``The Dots are Connected,'' Proc. OFC, PDP41, San Diego (2010).
%\bibitem{ref6}
%F. M. Lastname, How to Write a Paper, Publishing Company (2002).
%\end{thebibliography}
%}

%\end{spacing}
%\vspace{-4mm}

%%%%%%%%%%%%%%%%%%%%%%%%%%%%%%%%%%%%%%%%%%%%%
%---------------------------------------------- End of Document -----------------------------------------------%
\end{document}